\documentclass[debug,usenatbib]{rmaa}

\usepackage{paralist}
\usepackage{longtable}

\usepackage{psfrag,color}

\usepackage[latin1]{inputenc}

\title{Connecting the formation of stars and planets. II: coupling the angular momentum of stars with the angular momentum of planets} 

\author{
  L. M. Flor-Torres, \altaffilmark{1} 
  R. Coziol,\altaffilmark{1}
  K.-P. Schr\"oder,\altaffilmark{1}
  D. Jack,\altaffilmark{1}
  and J. H. M. M. Schmitt,\altaffilmark{2}
  }

\altaffiltext{1}{Departamento de Astronom\'{\i}a, Universidad de Guanajuato, Guanajuato, Gto, M\'exico.}
\altaffiltext{2}{Hamburger Sternwarte, Universit\"at Hamburg, Hamburg, Germany.}

\shortauthor{Flor-Torres et al.}
\shorttitle{RevMexAA Main Journal Demo Document}

\fulladdresses{
\item L. M. Flor-Torres, R. Coziol, K.-P. Schr\"oder, D. Jack : Departamento de Astronom\'{\i}a, Universidad de Guanajuato, Guanajuato, Gto, M\'exico.
\item J. H. M. M. Schmitt: Hamburger Sternwarte, Universit\"at Hamburg, Hamburg, Germany.}
\indexauthor{Flor-Torres, L. M.}
\indexauthor{Coziol, R.}
\indexauthor{Schr\"oder, K.-P.}
\indexauthor{Jack, D.}
\indexauthor{Schmitt, J. H. M. M.}

\abstract{
A sample of 46 stars, host of exoplanets, is used to search for a connection between their formation process and the formation of the planets rotating around them. Separating our sample in two, stars hosting high-mass exoplanets (HMEs) and low-mass exoplanets (LMEs), we found the former to be more massive and to rotate faster than the latter. We also found the HMEs to have higher orbital angular momentum than the LMEs and to have lost more angular momentum through migration. These results are consistent with the view that the more massive the star and higher its rotation, the more massive was its protoplanetarys disk and rotation, and the more efficient the extraction of angular momentum from the planets. 

}

\resumen{
Una muestra de 46 estrellas, hu\'espedes de exoplanetas, esta usada para la b\'usqueda de una conexi\'on entre su proceso de formaci\'on y el proceso de formaci\'on de los planetas que las orbitan. Separando nuestra muestra en dos, estrellas hu\'espedes de exoplanetas de baja masa (LME) y de alta masa (HME), encontramos que las HMEs rotan preferiblemente alrededor de estrellas con alta masa y con mayor rotaci\'on que las estrellas de las LMEs. Tambi\'en se encontr\'o que las HMEs tienen un momento angular orbital m\'as alto que las LMEs y que perdieron m\'as momento angular durante su migracion. Nuestros resultados son consistente con un modelo donde estrellas m\'as masivas con alta rotacion forman discos protoplanetarios m\'as masivos que tambien rotan mas r\'apido, y que ademas son mas eficientes en disipar el momento angular de sus planetas. 
}

\addkeyword{stars: fundamental parameters}
\addkeyword{stars: rotation}
\addkeyword{stars: formation}
\addkeyword{stars: planetary systems}

\begin{document}
\maketitle

\section{Introduction}
\label{sec:intro}

The discovery of gas giant planets rotating very close to theirs stars (hot Jupiter, or HJs) has forced us to reconsider our model for the formation of planets around low mass stars by including in an ad hoc way large scale migration. Since this did not happen in the solar system, it brings the natural question of understanding under what conditions large scale migration could be triggered in a proto-planetary disk (PPD). By stating such question, we adopt the simplest view that there is only one universal process for the formation of planets, following the collapse of dense regions in a molecuar cloud \citep{McKee2007,Draine2011,Champion2019}. This reduces the problem to a more specific one which is: how do we include migration in a well developed model like the core collapse scenario (the standard model), which explains in details how the solar system formed \citep{Safronov1969,Wetherill1989,Wurm1996,Poppe1997,Klahr2006,Hilke2011,dePater2015,Raymond2020}?  

In the literature, two migration mechanisms are favored for HJs \citep{Dawson2018,Raymond2020}. The first is disk migration \citep[e.g., ][and references therein]{Baruteau2014, Armitage2020}, which proposes that a planet looses its orbit angular momentum by tidal interactions with the PPD, while the second, high-eccentricity migration \citep[e.g.,][]{Rasio1996,Weidenschilling1996,Marzari2002,Chatterjee2008,Nagasawa2008,Beauge2012}, suggests a planet interacting with other planets gains a higher eccentricity, which brings it close to its star where it reaches equilibrium by tidal interactions (a process know as circularization). In terms of migration, these two mechanisms might suggest massive disks  somehow amplified the level of migration compared to what happened in the solar system, because more massive PPDs either increase the intensity of interactions of the planets with their disks or favor the formation of a higher number of planets. Within the standard model this would suggest that what counts is whether the PPD follows the minimum mass model, with a mass between 0.01 to 0.02 M$_\odot$, or the maximum mass model with a mass above 0.5 M$_\odot$ \citep{Armitage2010}. There are few clues which could help us determining which path the PPD of the solar system followed \citep[and strong difficulties compared to direct observations of PPD; see Fig. 2 and discussion in][]{Raymond2020}. One is the total mass of the planets, which represents only 0.1\% the mass of the Sun. This implies the solar system PPD have lost an important amount of its mass after the formation of the planets. Another clue is that 99\% of the angular momentum of the solar system is located in the planets, suggesting that the initial angular momentum of the PPD might have been conserved in the planets. However, this is obviously not the case when large scale migration occurs, so what was the difference?

If the initial angular momentum of the PPD passes to the planets, then one could use the orbital rotation momentum in exoplanetary systems to test different scenarios connecting the formation of the planets to the formation of their stars.  For example, how is the angular momentum of the PPD coupled to the angular momentum of the stars? Since large scale migration represents a loss of angular momentum of the planets (at least by a factor 10), what was the initial angular momentum of the PPD when it formed and how does this compared to the initial mass of the PPD? Does this influence the masses of the planets and their migration?The answers are not trivial, considering that the physics involved is still not fully understood. 

In particular, we know that the angular momentum is not conserved during the formation of stars. This is obvious when one compares how fast the Sun rotates with how fast its rotation should have been assuming the angular momentum of the collapsing molecular cloud where it formed was conserved. Actually, working the math \citep[a basic problem, but quite instructive; see course notes by][]{Alexander2017}, the Sun effective angular momentum, $j_\odot = J_\odot/M_\odot$, is $\sim 10^6$ times lower than expected. Intriguingly, $j_\odot$ is also $10^3$ lower than the angular momentum of its breaking point, $j_b$, the point where the centripetal force becomes stronger than gravity \citep{McKee2007}. If that was not true, then no stars whatever massive would be able to form. In fact, observations revealed that, in general, the angular momentum of stars with spectral type O5 to A5 trace a power law, $J \propto M^\alpha$, with $\alpha \sim 2$, with typical $J_{*}$ values that are exactly ten times lower than their breaking point. How universal is this ``law''and how stars with different masses get to it, however, is unexplained \citep{Wesson1979,Brosche1980,Carrasco1982,Godlowski2003}. To complicate the matter, it is clear now that lower mass stars, later than A5, do not follow this law, their spin going down exponentially (cf. Fig.6 in Paper~{\rm I}). For low-mass stars, \citet{McNally1965}, \citet{Kraft1967} and \citet{Kawaler1987} suggested a steeper power law, $J \propto M^{5.7}$, which suggests they loose an extra amount of angular momentum as their mass goes down. What is interesting is that low-mass stars are also those that form PPDs and planets, which had led some researchers to speculate there could be a link between the two. 

To explain how low-mass stars loose their angular momentum, different mechanisms are considered. The most probable is stellar wind \citep{Schatzman1962,Mestel1968}, which is related to the convective envelopes of these stars. This is how low-mass stars would differ from massive ones. However, whether this mechanism is sufficient to explain the break in the $J-M$ relation is not obvious, because it ignores the possible influence of the PPD \citep[the formation of a PPD seems crucial; see][]{delaReza2004}. This is what the magnetic braking model takes into account \citep{Wolff1997, Tassoul2000,Uzdensky2002}. Being bombarded by cosmic rays and UV radiation from ambient stars, the matter in a molecular cloud is not neutral, and thus permeable to magnetic fields. This allows ambipolar diffusion (the separation of negative and positive charges) to reduce the magnetic flux, allowing the cloud to collapse. Consequently, a diluted field follows the matter through the accretion disk to the star forming its magnetic field  \citep{McKee2007}. This also implies that the accretion disk (or PPD) stays connected to the star through its magnetic field as long as it exists, that is, a period that although brief includes the complete phase of planet formation and migration. According to the model of disk-locking, a gap opens between the star and the disk at a distance $R_t$ from the star, and matter falling between $R_t$ and the radius of corotation, $R_{co}$ (where the Keplerian angular rotation rate of the PPD equals that of the star), follow the magnetic field to the poles of the star creating a jet that transports the angular momentum out. In particular, this mechanism was shown to explain why the classic T-Tauri rotates more slowly than the weak T-Tauri \citep{Ray2012}. How this magnetic coupling could influence the planets and their migrations, on the other hand, is still an open question \citep{Matt2004,Fleck2008,Champion2019}.

To investigate further these problems, we started a new observational project to observe host stars of exoplanets using the 1.2 m robotic telescope TIGRE, which is installed near our department at the La Luz Observatory (in central Mexico). In paper~{\rm I} we explained how we succeeded in determining in an effective and homogeneous manner the physical characteristics ( $T_{eff}$,  $\log g$, [M/H], [Fe/H], and  $V \sin i$) of a initial sample of 46 bright stars using \textsf{iSpec} \citep{Blanco-Cuaresma2014,Blanco-Cuaresma2019}. In this accompanying article, we now explore the possible links between the physical characteristics of these 46 stars and the physical characteristics of their planets, in order to gain new insight about a connection between the formation of stars and their planets.

\section{Sample of exoplanets}
\label{sample}

Our observed sample consists of 46 stars host of 59 exoplanets, which were selected from the revised compendium of confirmed exoplanets in the Exoplanet Orbit Database\footnote{http: //exoplanets.org/}. In Table~\ref{tab:0} the dominant planet (col.~2) in each stellar system is identified by the same number (col.~1) which was used in paper~I to identify their stars. In col.~3 and 4 we repeat the magnitude and distance of the host stars as they appeared in Table~1 of paper~I. This is followed by the main properties of the planets as  reported in the exoplanet Orbit Database: mass (col.~5), radius (col.~6), period (col.~7), major semi-axis (col.~8) and eccentricity (col.~9). Note that an eccentricity of zero could mean the actual eccentricity is not known.  The last two columns identify the detection method and the  distinction between high mass and low mass exoplanet (as explained below). 

Among our sample, one exoplanet, \#~39 (Hn-Peg b), was found to have a mass above the Brown Dwarf (BD) lower limit of $13\ M_{Jup}$ \citep[][]{Spiegel2011,Burgasser2008}. Note that Hn-Peg b was detected in imaging (identified as Im in col.~11), and its huge distance from its host star (col.~8) is more typical of BDs than of exoplanets (the other exoplanet detected in imaging is 34, HD~115383 b, at 43.5 AU from it stars, but with a mass of only 4 $M_{Jup}$). Another exoplanet whose mass is close to the lower mass limit for BDs is \#~31 (XO-3 b), but since it is located very near its star, 0.05 AU, there is no difficulty in accepting its classification as exoplanet. Due to the small size of our sample, we cannot assess how the formation of exoplanets with masses above the BD limit can differ from the formation of the other exoplanets. Consequently, although we kept Hn-Peg in our sample of host stars we did not included its ``exoplanet'' in our statistical analysis.

\begin{table*}
 \begin{minipage}{\textwidth}
 \centering
 \caption[]{\small{Physical parameters the planets in our sample.}}
 \label{tab:0}
  \vspace{0.25cm} 
 \begin{Huge}
 \resizebox{\textwidth}{!}{%
\begin{tabular}{clccccccccc}
\hline
\textbf{Id. \#} 	& \textbf{Star and} 	&	\textbf{Magnitude}	&	\textbf{Distance}	& \textbf{$M_p$} 	& \textbf{$R_p$} 	& \textbf{Period} 	& \textbf{$a_p$} 	& \textbf{$e_p$}  	&	\textbf{Detection}	& \textbf{Planetary} 	\\
 	&  Planet	&	(V)	&	(pc)	& (M$_{jup}$) 	& (R$_{jup}$) 	& (days) 	& (AU )	&  	&	\textbf{Method}	&  \textbf{type}	\\
\hline
\hline
1	    & *KELT-6 c 	    &	10.3		&	242.4	&	3.71		&	2.68	&	1276.0		&	2.39	&	0.21	&	RV 	& HME 	\\
2	    & *HD 219134 h &	5.6   	&	6.5		&	0.28		&	0.80	&	2198.0		&	3.06	&	0.37	&	RV	& LME 	\\
3	    & *KEPLER-37 b 	&	9.8  		&	64.0		&	0.01		&	0.03	&	13.37		&	0.10	&	0		&	Tr		& LME	\\
4	    & HD 46375 b 	&	7.8		&	29.6		&	0.23		&	1.02	&	3.02			&	0.04	&	0.05	&	RV	& LME 	\\
5	    & HD 75289 b 	&	6.4		&	29.1		&	0.47		&	1.03	&	3.51			&	0.05	&	0.02	&	RV	& LME 	\\
6	    & HD 88133 b 	&	8.0		&	73.8		&	0.30		&	1.00	&	3.42			&	0.05	&	0.08	&	RV	& LME 	\\
7	    & HD 149143 b 	&	7.9		&	73.4		&	1.33		&	1.05	&	4.07			&	0.05	&	0.01	&	RV	& HME 	\\
8	    & HAT-P-30 b 	   	&	10.4		&	215.3	&	0.71		&	1.34	&	2.81			&	0.04	&	0.04	&	Tr		& LME 	\\
9	    & KELT-3 b 	       	&	9.8		&	211.3	&	1.42		&	1.33	&	2.70			&	0.04	&	0		&	Tr		& HME 	\\
10	& KEPLER-21 b 	&	8.3		&	108.9	&	0.02		&	0.15	&	2.79			&	0.04	&	0.02	&	Tr		& LME 	\\
11	& KELT-2A b 	   	&	8.7		&	134.6	&	1.49		&	1.31	&	4.11			&	0.05	&	0.19	&	Tr		& HME 	\\
12	& HD86081 b 	    &	8.7		&	104.2	&	1.50		&	1.08	&	2.00			&	0.04	&	0.06	&	RV	& HME 	\\
13	& WASP-74 b 	    &	9.8		&	149.8	&	0.97		&	1.56	&	2.14			&	0.04	&	0		&	Tr		& LME 	\\
14	& HD 149026 b 	&	8.1		&	76.0		&	0.36		&	0.72	&	2.88			&	0.04	&	0		&	Tr		& LME 	\\
15	& HD 209458 b 	&	7.6		&	48.4		&	0.69		&	1.38	&	3.52			&	0.05	&	0.01	&	Tr		& LME 	\\
16	& BD-10 3166 b 	&	10.0		&	84.6		&	0.46		&	1.03	&	3.49			&	0.05	&	0.01	&	RV	& LME 	\\
17	& HD 189733 b 	&	7.6		&	19.8		&	1.14		&	1.14	&	2.22			&	0.03	&	0		&	Tr		& LME 	\\
18	& HD 97658 b 	&	7.7		&	21.6		&	0.02		&	0.20	&	9.49			&	0.08	&	0.08	&	Tr		& LME 	\\
19	& HAT-P-7 b 	    &	10.5		&	344.5	&	1.74		&	1.43	&	2.20			&	0.04	&	0		&	Tr		& HME 	\\
20	& KELT-7 b 	        &	8.5		&	137.2	&	1.29		&	1.53	&	2.73			&	0.04	&	0		&	Tr		& HME 	\\
21	& HAT-P-14 b     	&	10.0		&	224.1	&	2.20		&	1.20	&	4.63			&	0.06	&	0.10	&	Tr		& HME 	\\
22	& WASP-14 b 	    &	9.7		&	162.8	&	7.34		&	1.28	&	2.24			&	0.04	&	0.09	&	Tr		& HME 	\\
23	& HAT-P-2 b 	    &	8.7		&	128.2	&	8.74		&	0.95	&	5.63			&	0.07	&	0.52	&	Tr		& HME 	\\
24	& WASP-38 b 	    &	9.4		&	136.8	&	2.71		&	1.08	&	6.87			&	0.08	&	0.03	&	Tr		& HME 	\\
25	& HD 118203 b 	&	8.1		&	92.5		&	2.14		&	1.05	&	6.13			&	0.07	&	0.29	&	RV	& HME 	\\
26	& HD 2638 b 	    &	9.4		&	55.0		&	0.48		&	1.04	&	3.44			&	0.04	&	0.04	&	RV	& LME 	\\
27	& WASP-13 b 	    &	10.4		&	229.0	&	0.49		&	1.37	&	4.35			&	0.05	&	0		&	Tr		& LME 	\\
28	& WASP-34 b 	    &	10.3		&	132.6	&	0.59		&	1.22	&	4.32			&	0.05	&	0.04	&	Tr		& LME 	\\
29	& WASP-82 b 	    &	10.1		&	277.8	&	1.24		&	1.67	&	2.71			&	0.04	&	0		&	Tr		& HME 	\\
30	& HD 17156 b 	&	8.2		&	78.3		&	3.20		&	1.10	&	21.22		&	0.16	&	0.68	&	Tr		& HME 	\\
31	& XO-3 b 				&	9.9		&	214.3   &	11.79	&	1.22	&	3.19			&	0.05	&	0.26	&	Tr		& HME 	\\
32	& HD 33283 b 	&	8.0		&	90.1		&	0.33		&	0.99	&	18.18		&	0.17	&	0.46	&	RV	& LME 	\\
33	& HD 217014 b 	&	5.5		&	15.5		&	0.47		&	1.90	&	4.23			&	0.05	&	0.01	&	RV	& LME 	\\
34	& HD 115383 b 	&	5.2	   	&	17.5		&	4.00		&	0.96	& $-$ 			&	43.5	&	0		&	Im		& HME 	\\
35	& HAT-P-6 b 	   	&	10.5		&	277.5	&	1.06		&	1.33	&	3.85			&	0.05	&	0		&	Tr		& LME 	\\
36	& *HD 75732 d	&	6.0	    &	12.6		&	3.86		&	2.74	&	4867.0		&	5.45	&	0.03	&	RV	& HME 	\\
37	& HD 120136 b	&	4.5	    &	15.7		&	5.84		&	1.06	&	3.31			&	0.05	&	0.08	&	RV	& HME 	\\
38	& WASP-76 b		&	9.5	  	&	195.3	&	0.92		&	1.83	&	1.81			&	0.03	&	0		&	Tr		& LME 	\\
39	& Hn-Peg b 	       &	6.0		&	18.1		&	16.00	&	1.10	& $-$ 			&	795 &	0		&	Im		& BD 	\\
40 	& WASP-8 b 		& 9.9 		& 90.2 	& 2.24 	& 1.04 & 8.16 		& 0.08 & 0.31 & Tr 		& HME 	\\
41 	& WASP-69 b 		& 9.9  		& 50.0 	& 0.26 	& 1.06 & 3.87 		& 0.05 & 0.00 & Tr 		& LME 	\\
42 	& HAT-P-34 b 		& 10.4  	& 251.1 	& 3.33 	& 1.11 & 5.45 		& 0.07 & 0.44 & Tr		& HME 	\\
43 	& HAT-P-1 b 		& 9.9  		& 159.7 	& 0.53 	& 1.32 & 4.47 		& 0.06 & 0.00 & Tr		& LME 	 \\
44 	& WASP-94 A b 	& 10.1  	& 212.5 	& 0.45 	& 1.72 & 3.95 		& 0.06 & 0.00 & Tr		& LME 	\\
45	& WASP-111 b 	& 10.3  	& 300.5 	& 1.85 	& 1.44 & 2.31 		& 0.04 & 0.00 & Tr 		& HME 	\\
46 	& HAT-P-8 b 		& 10.4  	& 212.8 	& 1.34 	& 1.50 & 3.08 		& 0.04 & 0.00 & Tr 		& HME 	\\
\hline
\multicolumn{11}{l}{An * in front of the name of the planet identified multiple planetary systems.}
 \end{tabular}
 }
 \end{Huge}
 \end{minipage}
 \end{table*}

Another point of importance that can be noted in Table~\ref{tab:0} is the fact that we have only 14 exoplanets detected in RV. This is most probably due to the fact that we required the radius of the planets to be known, which is easier to determine by the Tr method. Curiously, only three of the 14 exoplanets detected in RV show the trend to be located farther from their stars than the Tr exoplanets, as is observed in the literature (and as is obvious in the Exoplanet Orbit Database). Two are close to the ice line and thus are more warm than hot, and one with 3.86 $M_{Jup}$ is at the same distance as Jupiter from the Sun. This implies that any bias introduced by the different detection methods, RV vs. Tr, cannot be explored thoroughly in our present analysis.  

Although the variety of the characteristics of exoplanets cannot be addressed with our present sample, we can however separate our sample of exoplanets in two based on their masses. For our analysis, this distinction is important in order to test how the mass of the exoplanet is related to the mass of its host star. To use a mass limit that has a physical meaning we choose $1.2\ M_{Jup}$, which is the mass above which self-gravity in a planet becomes stronger than the electromagnetic interactions \citep[][see demonstration in Appendix~\ref{sec:ap-A}]{Padmanabhan1993,Flor-Torres2016}. Using this limit we separated our sample in 22 high mass exoplanets (HMEs) and 23 low mass exoplanets (LMEs). This classification is included in Table~\ref{tab:0} in the last column. 

According to \citet{Fortney2007} the mass-radius relation of exoplanets show a trend for exoplanets above  $1.0\ M_{Jup}$ to have a constant radius \citep[see also][]{Fortney2010}. In fact, modelizations of exoplanet structures \citep[e.g.,][]{Baraffe1998,Baraffe2003,Baraffe2008} predicts an inflection point in the mass-radius relation where the radius starts to decrease instead of increasing as the mass increases. Actually, this is what we observe in brown dwarfs (BDs). Physically therefore, this inflection should be located near $1.2\ M_{Jup}$, where self-gravity becomes stronger than the electromagnetic interaction and the object start to collapse (which is the case of BDs). However, if massive HJs have more massive envelopes of liquid metallic hydrogen (LMH) than observed in Jupiter \citep[as suggested by JUNO;][]{Guillot2018,Kaspi2018,Iess2018,Adriani2018}, their structures might resist gravity (at least for a while), due to the liquid state being incompressible, pushing the collapse of the radius to slightly higher masses \citep[models and observations verifying this prediction can be found in][]{Hubbard1997a,Dalladay-Simpson2016,Flor-Torres2016}. This possibility, however, is still controversial, and we only use the mass limit in our analysis to separate our sample of exoplanets according to their masses. On the other hand, the possibility of massive LMH envelopes in the HMEs might have some importance, since these exoplanets would be expected to have higher magnetic fields than the LMEs , which, consequently, could affect their interactions with the PPD and nearby host stars \citep[this fits the case of HD~80606~b, a HME studied by][]{deWit2016}, leading possibly to different migration behaviors. Likewise, we might also consider the possibility of inflated radii in the LMEs, since they are so close to their stars, although what could be the effect of inflated radius on migration is less obvious (one possibility is that they circularize more rapidly). 

\begin{figure}[h!]
\includegraphics[width=0.8\linewidth, angle=270]{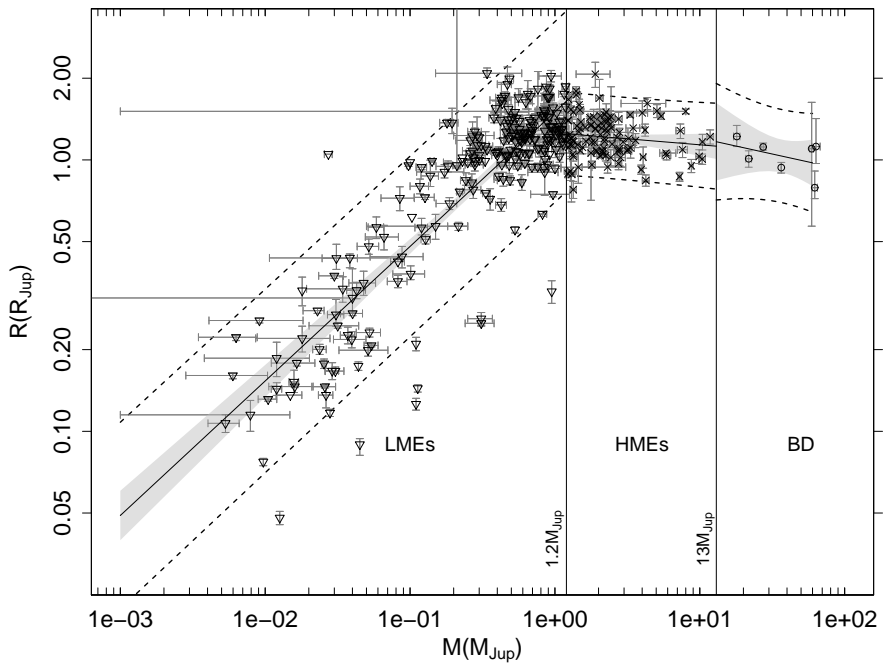}
\caption{The M-R diagram for exoplanets in the Exoplanet Orbit Database. The vertical line at $1.2\ M_{Jup}$ corresponds to the mass criterion we used to separate the exoplanets in LMEs and HMEs. The other vertical lime is the lower mass limit for the BDs, $13\ M_{Jup}$.}
\label{f1}
\end{figure}

To test further our distinction in mass, we trace in Fig.~\ref{f1} the mass-radius relation of 346 exoplanets from the Exoplanet Orbit Database. Based on visual inspection it is not clear due to inflated radii how to separate the HJs. One quantitative criterion is the mass-radius relation. In figure \ref{f1}, we traced three M-R relations, adopting $1.2\ M_{Jup}$ to distinguish LMEs form HMEs. Below this limit we find the relation:
\begin{equation}\label{eq1}
\ln R_{LME} = (0.496\pm 0.018) \ln M_{\rm LME} + (0.407 \pm 0.034)
\end{equation}
The slope is positive and the correlation coefficient is high, $r^2 = 0.75$, which implies that the radius within this range of masses continually increases with the mass. This relation is also fully consistent with what was previously reported by \citet{Valencia2006} and \citet{Chen2017}. Above the lower mass limit for BDs we find the relation: 
\begin{equation}\label{eq3}
\ln R_{BD} = (-0.117\pm 0.111) \ln M_{\rm BD} + (0.455 \pm 0.404)
\end{equation}
with a negative slope and a weaker correlation coefficient of $r^2 = 0.18$, but which is sufficient to indicate that the trend is for the radius to decrease with the mass.  This is as expected for objects where self-gravity is stronger than the electromagnetic repulsion (BDs not having enough mass to ignite fusion in their core cannot avoid the effect of gravitational collapse). Finally, in between these two mass limits defining the HMEs, we obtain the relation: 
\begin{equation}\label{eq2}
\ln R_{HME} = (-0.044\pm 0.030) \ln M_{\rm HME} + (0.229 \pm 0.031)
\end{equation}
which has an almost nil slope and a very weak coefficient of correlation, $r^2 = 0.02$, consistent with no correlation. This range of mass, therefore, is fully consistent with HJs near the inflection point, extending this region over a decade in mass \citep[as previously noted by][]{Hatzes2015}. For our analysis, these different M-R relations are sufficient to justify our separation between LMEs and HMEs \citep[note that this physical distinction criterion was never used previously in the literature; the only study which uses a limit close to ours is][]{Sousa2011}.

\section{Connecting the stars to their planets}

 \begin{figure}
\hspace{-0.8cm}
\includegraphics[width=0.7\linewidth, angle=270]{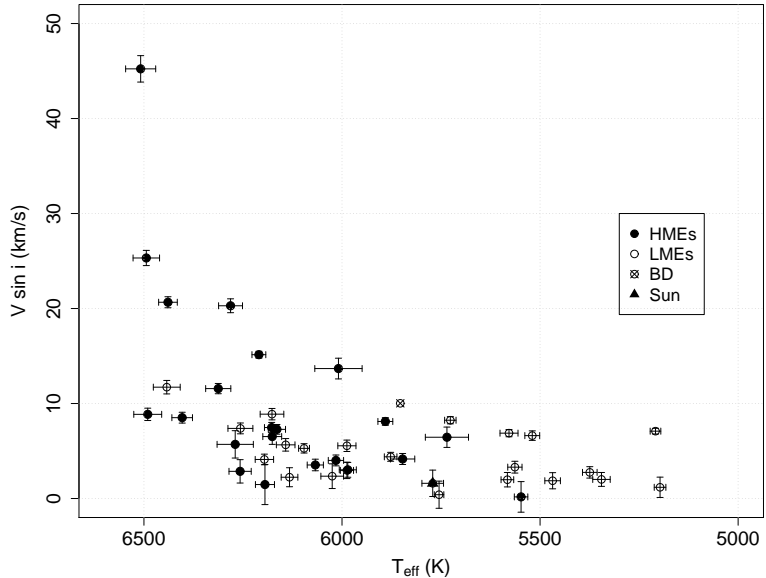}
\caption{\textit{Star rotational velocity vs. temperature, distinguishing between stars hosting HMEs and LMEs. The position of the Sun is included as well as the star with a BD as companion.}}
\label{f2}
\end{figure}

\begin{figure}
\includegraphics[width=1.0\linewidth, angle=0]{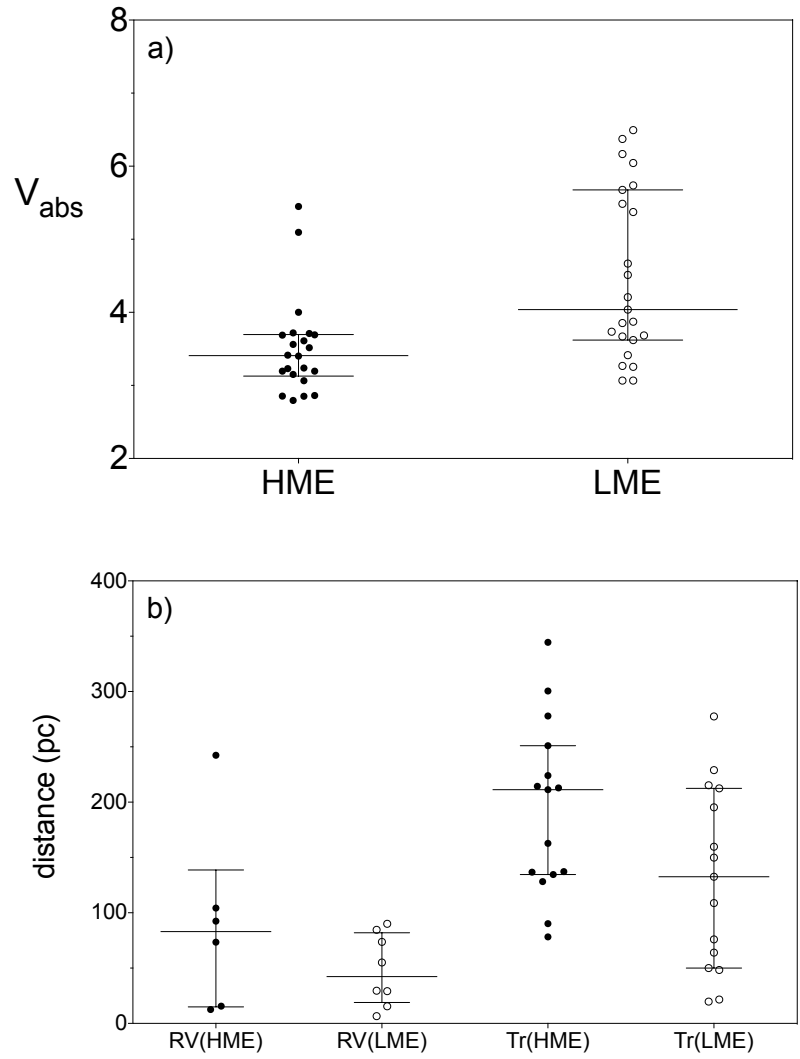}
\caption{a) Absolute magnitude in V distinguishing stars hosting HMEs and LMEs; b) Distance of the stars also distinguishing by the detection method, radial velocity, RV, or transit, Tr. The bars correspond to the medians and interquartile ranges.}
\label{f3}
\end{figure}

In paper~{\rm I}, we verified that the rotational velocity of a star, $V \sin i$, decreases with the temperature, $T_{eff}$. In Fig.~\ref{f2}, we reproduced this graphic, but this time distinguishing between stars hosting LMEs and HMEs. We observe a clear trend for the HMEs to be found around hotter and faster rotator stars than the LMEs. Considering the small number of stars in our sample, we need to check whether this result is physical or due to an observational bias. For example, one could suggest that HMEs are easier to detect than LMEs through the RV than Tr method around hotter (more massive) and faster rotator stars. To check for observational biases, we trace first in Fig.~\ref{f3}a the distributions of the absolute V magnitude for the stars hosting LMEs and HMEs. We do find a trend for the stars hosting HMEs to be more luminous than the stars hosting LMEs. This is confirmed by a non-parametric Mann-Whitney test with a p-value of 0.0007 \citep{Dalgaard2008}. However, in Fig.~\ref{f3}b we also show that the reason why the stars with HMEs are more luminous is because they are located farther out, and this is independent of the detection method. Consequently, the trend for the HMEs to be found around hotter and faster rotator stars than the LMEs does not depend on the method of detection, but is a real physical difference. Considering the mass-temperature relation on the main sequence, this suggests that the more massive exoplanets in our sample rotate around more massive stars. 

\begin{figure}
\hspace{-0.8cm}
\includegraphics[width=1.2\linewidth, angle=0]{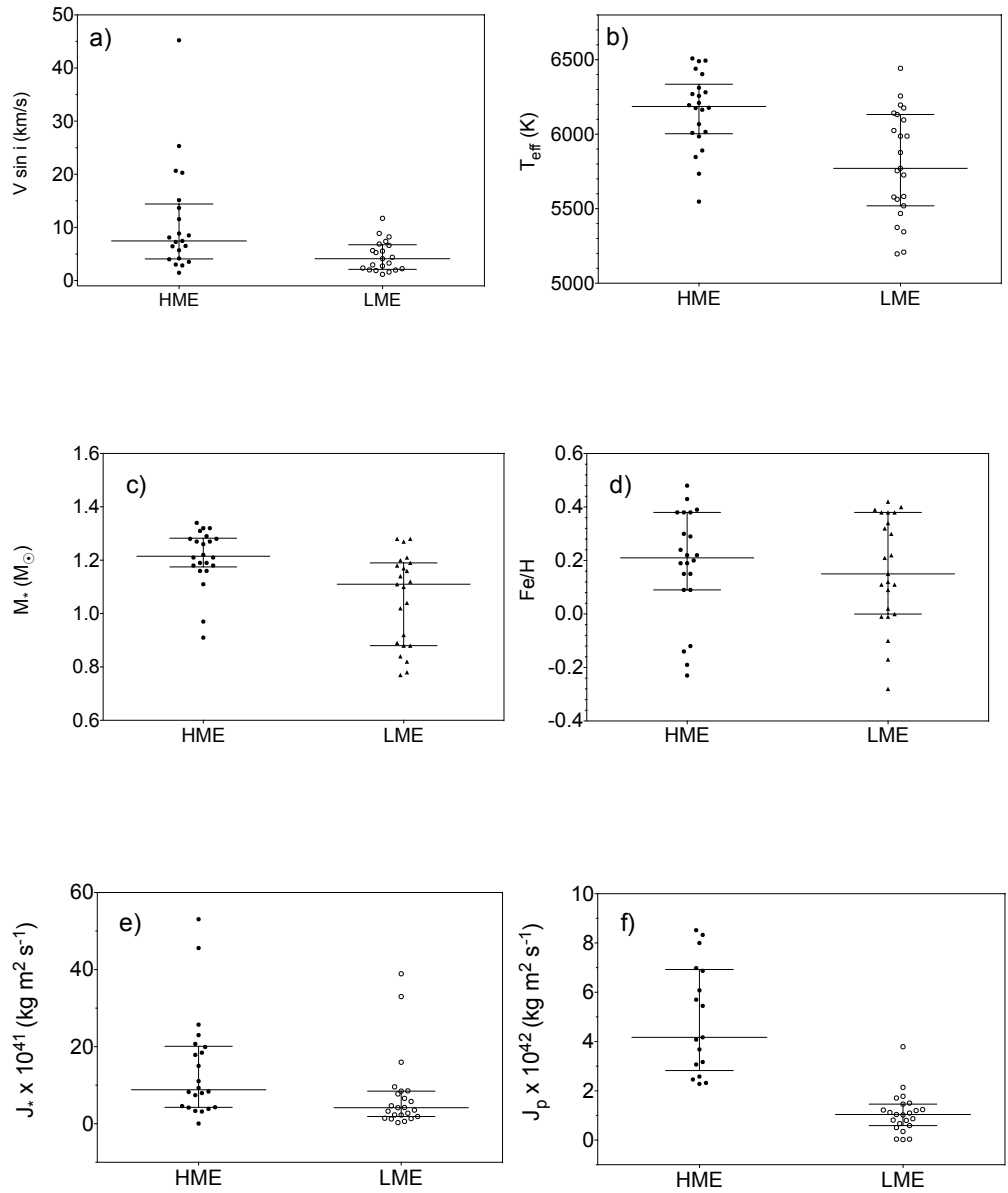}
\caption{\textit{Comparing the physical characteristics of the stars and planet hosting HMEs and LMEs. In each graph the medians and interquartile ranges are drawn over the data. a) rotational velocity, b) effective temperature, c) mass of the star, d) metallicity, e) the angular momentum of the star, f) the angular momentum of the orbit of the dominant planet.}}
\label{f4}
\end{figure}

In Fig.~\ref{f4} we compare the physical characteristics of the stars that host HMEs with those that host LMEs. Fig.~\ref{f4}a shows a clear trend for rotational velocity to be higher in the stars hosting HMEs than LMEs. In Table~\ref{tab:5} we give the results of non-parametric, Mann-Whitney (MW) tests. The use of non-parametric tests is justified by the fact that we did not find normal distributions for our data (established by running 3 different normality tests). In col.~6 we give the p-values of the tests at a level of confidence of 95\%, in col.~7, the significance levels (low, *, medium, ** and high, ***) and in col.~8 the acceptance or not of the differences observed. Note that a non-parametric test compares the ranks of the data around the medians, not the means. In the case of $V \sin i$, the MW test confirms the difference of medians at a relatively high level of confidence (the stars 22, 33 and 36 were not considered due to their high uncertainties). In Fig.~\ref{f4}b we see the same trend for $T_{eff}$ on average, also with a significant difference in median in Table~\ref{tab:5}. 

\begin{table}
 \centering
 \caption[]{\small{Statistical tests}}
\label{tab:5}
  \vspace{0.25cm} 
 \begin{small}
\begin{tabular}{cccccccccc}
      \toprule   
       & \multicolumn{2}{c}{HME} &  &\multicolumn{2}{c}{LME} & & MW  & sign. & diff. \\
      \cmidrule(l){2-3}\cmidrule(l){5-6}\\
       Parameter  & median & mean & & median & mean & & p-value &  & \\
      \midrule
 $V \sin i$		&	$7.47$	   &	$10.94$ 	&&	$4.11$		& $4.62$			&&$0.0048$		&	**    & yes \\	
 $T_{eff}$		&	$6186$	&	$6158$ 	&&	$5771$	& $5800$		&&$< 0.0001$	&	*** 	& yes \\
 $M_{*}$	    	&	$1.22	$	&	$1.21	$ 	&&	$1.11$		& $1.05$			&&$< 0.0001$	&	***	& yes  \\
 $[Fe/H]$	 		&	$0.21$		&	$0.19$ 	&&	$0.15$		&$0.16$  		&&$0.6810$		&	ns		&	no   \\
 $J_{*}$		 	&	$8.83	$	&	$14.3$ 	&&	$4.17$		&$7.33$			&&$0.0096$		&	** 	& yes  \\
 $J_{p}$		 	&	$4.17$		&	$4.93$ 	&&	$1.04$	   	&$1.09$		   	&&$< 0.0001$	&	*** 	& yes  \\
\hline
\multicolumn{10}{l}{The units and scale values for the medians and means are those of Fig.~\ref{f4}}
\end{tabular} 
\end{small}
 \end{table}
 
In Figure~\ref{f4}c the difference in mass is obvious, and confirmed by the MW test at the highest level of confidence. Stars with HMEs are more massive than stars with LMEs. On the other hand we distinguish no difference in the distribution of the metallicity.  Although the medians and means reported in Table~\ref{tab:5} look different, the p-value of the MW test is unequivocal, being much higher than 0.05. The median [Fe/H] for our sample is 0.07 dex and the mean is 0.12 dex, with a standard deviation of 0.24 dex. These values are comparable to what is reported in the literature for systems harboring massive HJ exoplanets \citep{Sousa2011,Buchhave2012}.  

In figure  Fig.~\ref{f4}d and 4f, once again we distinguish obvious differences in the distributions, the stars with HMEs having higher angular momentum than the stars with LMEs, and the HMEs having also higher angular momentum than the LMEs. The MW tests in Table~\ref{tab:5} confirmed these differences at a relatively high level of confidence (like for $V \sin i$). Considering Eq.~2 in paper~{\rm I} for the angular momentum of the star, the statistical test confirms that the HMEs rotate around more massive and faster rotator stars than the LMEs. Considering the difference in angular momentum of the exoplanets, $J_p$, this result seems to support the hypothesis that more massive planets form in more massive PPDs with higher angular momentum. The question, then, is how does this difference affects the migration process of the different exoplanets?

\begin{figure*}
\includegraphics[width=0.8\linewidth, angle=270]{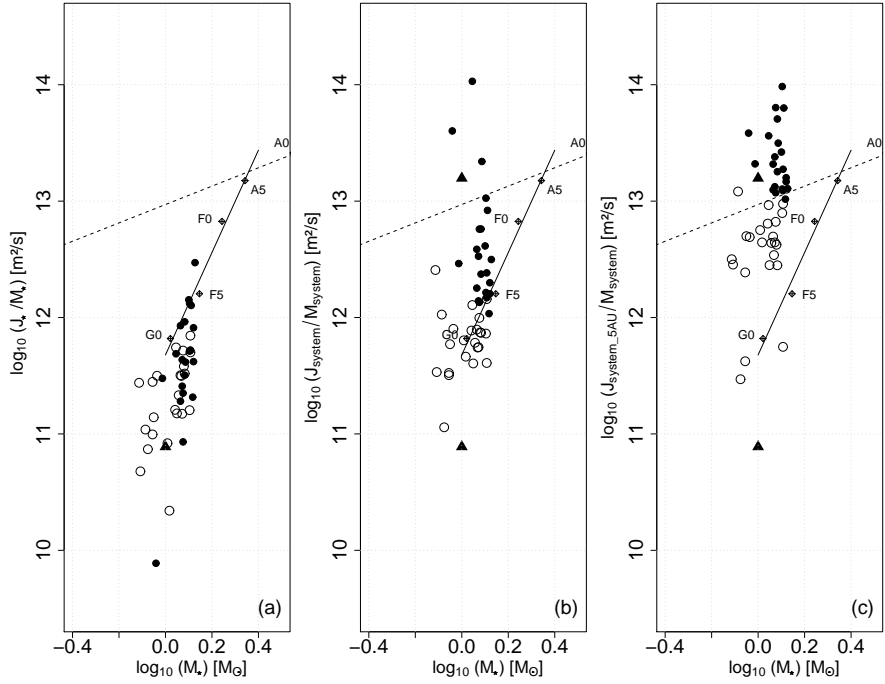}
\caption{\textit{In (a) Specific angular momentum of the host stars as function of their masses. The symbols are as in Fig.~\ref{f2}. The solid black line is the relation proposed by \citet{McNally1965} for low-mass stars (spectral types A5 to G0) and the dotted line is the extension of the relation suggested for massive stars. The black triangle represents the Sun; (b) specific angular momentum for the planetary systems, with the inverted black triangle representing the Solar system; (c) original angular momentum assuming the planets formed at 5 AU.}}
\label{f5}
\end{figure*}

In Fig.~\ref{f5}a, following \citet{Berget2010}, we trace the specific angular momentum of the stars, $j_{*} = J_{*}/M_{*}$, as a function of their masses, distinguishing between stars hosting HMEs and LMEs. The angular momentum of all the host stars in our sample fall well below the theoretical relation proposed by \citet{McNally1965}. There are is clear distinction between the LMEs and HMEs, except for the differences encountered in Fig.~\ref{f4} and confirmed in Table~\ref{tab:5}, and there is no evidence the stars follow a $j-M$ relation. 

On the other hand, when we compare in Figure~\ref{f5}b the angular momentum for the systems, $j_{sys} = j_{*} + j_{p}$, we do see a difference between systems with LMEs and HMEs. But this is expected, since, by definition, the HMEs having higher masses naturally have a higher contribution to $j_{sys}$. However, and despite being more massive than Jupiter, very few of the HME systems have a value of $j_{sys}$ comparable to the solar system. Obviously, this is because of the large scale migration they suffered. To illustrate this point, we traced in  Figure~\ref{f5}c the possible ``initial'' angular momentum the system could have had assuming the exoplanets formed at the same distance as Jupiter (5 AU). Comparing with the positions in Figure~\ref{f5}b, the HMEs would have lost on average 89\% of their initial momentum, compared to 86\% for the LMEs. Those losses are enormous. Considering the lost of angular momentum of the stars and planets, it might be consequently difficult to expect a coupling between $J_p$ and $J_{*}$ or even a $j-M$ relation. 

 \begin{table*}
 \begin{minipage}{\textwidth}
 \centering
 \caption[]{\small{Pearson correlation matrix for the HME systems}}
 \label{tab:3}
 \begin{Huge}
 \resizebox{\textwidth}{!}{%
 \begin{tabular}{cccccccccccc}
\hline
\textbf{HME}  & $T_{eff}$ 			& $\log g$ 				& $[Fe/H]$ 		& $V \sin i $ 			& $R_{*}$ 					& $M_{*}$  				& $J_{*}$ 	&  $M_{p}$ 				& $R_{p}$  		& $a_p$ 					& $e_p$ \\
\hline  
$\log g$ 			& 0.2918					& 								& 						& 								& 								& 								& 					& 								& 						& 								&			\\
$[Fe/H]$			& 0.6794					& 0.1519					& 						& 								& 								& 								& 					& 								& 						& 								&			\\
$V \sin i $		& {\bf 0.0048}		& 0.6052					& 0.4121			& 								& 								& 								& 					& 								& 						& 								& 				\\
$R_{*}$			& {\bf 0.0199}		& 0.0785					& 0.9143			& 0.8734					& 								& 								& 					& 								& 						& 								& 				\\
$M_{*}$			& $<${\bf 0.0001}	& 0.0884	 				& 0.4889			& 0.1242					& $<${\bf 0.0001}	& 								& 					& 								& 						& 								& 				\\
$J_{*}$			& {\bf 0.0006}		& 0.7412					& 0.1961			& $<${\bf 0.0001}	& 0.2345					& {\bf 0.0162}		& 					& 								& 						& 								& 				\\
$M_{p}$			& 0.8087					& 0.7356					& 0.3401			& 0.3785					& 0.0577					& 0.1892					& 0.7779	  	& 								& 						& 								&				 \\
$R_{p}$			& 0.3861					& 0.4608					& 0.2527			& 0.9065					& 0.2122					& 0.1460					& 0.8397		& 0.1323					& 						& 								&				 \\
$a_p$				& {\bf 0.0339}			& {\bf 0.0173}		& 0.6018			& 0.9447					& {\bf 0.0118}		& {\bf 0.0074}		& 0.3824		& {\bf 0.0074}		& 0.1050			& 								& 				\\
$e_p$				& 0.6455					& 0.4758					& 0.8288			& 0.9864 				& 0.4266					& 0.1871					& 0.8931		& {\bf 0.0073}		& {\bf 0.0418}& {\bf 0.0122}		& 				\\
$J_p$				& 0.4358					& 0.4193					& 0.4496			& 0.5536	 				& {\bf 0.0455}		& 0.5697					& 0.7913 	& $<${\bf 0.0001}	& 0.2679			& $<${\bf 0.0001}	& {\bf 0.0234	}\\    
\hline
$\log g$ 			& 								& 								& 						& 								& 								& 								& 					& 								& 						& 								&			\\
$[Fe/H]$			& 								& 								& 						& 								& 								& 								& 					& 								& 						& 								&			\\
$V \sin i $		& 0.5909					& 								& 						& 								& 								& 								& 					& 								& 						& 								& 				\\
$R_{*}$			& 0.4925 				& 								& 						& 								& 								& 								& 					& 								& 						& 								& 				\\
$M_{*}$			& 0.8402					&  							& 						& 								& 0.8662					&  							& 					& 								& 						& 								& 				\\
$J_{*}$			& 0.6736					& 								&						& 0.9857					& 								& 	0.5064					& 					& 								& 						& 								& 				\\
$M_{p}$			& 								& 								& 						& 								&  							& 								& 					& 								& 						& 								&				 \\
$R_{p}$			& 								& 								& 						& 								& 								& 								& 					& 								& 						& 								&				 \\
$a_p$				& 	-0.4539				& 0.5021					& 						& 								& -0.5267				& -0.5546				& 					& 0.5544					& 						& 								& 				\\
$e_p$				& 								& 								& 						& 								& 								&								& 					& 0.5550					& -0.4373		& 0.5243					& 				\\
$J_p$				& 								& 								& 						& 								&-0.4305 				& 								& 					& 0.9393					& 						& 0.7272					& 0.4811\\
 \hline
 \end{tabular}
 }
 \end{Huge}
 \end{minipage}
 \end{table*}
 
 \begin{table*}
 \begin{minipage}{\textwidth}
 \centering
 \caption[]{\small{Pearson correlation matrix for the LME systems}}
 \label{tab:4}
 \begin{Huge}
 \resizebox{\textwidth}{!}{%
 \begin{tabular}{cccccccccccc}
\hline
\textbf{LME}  & $T_{eff}$ 			& $\log g$ 				& $[Fe/H]$ 			& $V \sin i $ 				& $R_{*}$ 				& $M_{*}$  			& $J_{*}$ 	&  $M_{p}$ 				& $R_{p}$  			& $a_p$ 			& $e_p$ \\
\hline  
$\log g$ 		& {\bf 0.0002}		& 								& 							& 									& 								& 							& 					& 								& 							& 						&				\\
$[Fe/H]$		& 0.3461		& 0.2153					& 							& 									& 								& 							& 					& 								& 							& 						&				\\
$V \sin i $	& {\bf 0.0097}		& 0.0389					& 0.0761	 			& 									& 								& 							& 					& 								& 							& 						&          	\\
$R_{*}$		& $<${\bf 0.0001}	& $<${\bf 0.0001}	& 0.2374				& 	0.1499						& 								& 							& 					& 								& 							& 						& 				\\
$M_{*}$		& $<${\bf 0.0001}	& $<${\bf 0.0001}	& 0.0618				& {\bf 0.0187}			& $<${\bf 0.0001}	& 							& 					& 								& 							& 						& 				\\
$J_{*}$		& {\bf 0.0034}		& {\bf 0.0012}		& 0.1880 			& $<${\bf 0.0001}		& {\bf 0.0077}		& {\bf 0.0012}	& 					& 								& 							& 						& 				\\
$M_{p}$		& 0.0658					& 0.1672					& 0.5022				& 0.3259						& 0.3008 				& 0.1802				& 0.3683		& 								& 							& 						&			 	\\
$R_{p}$		& 0.1825					& 0.0681					& 0.5140				& 0.7583						& 0.1878					& 0.1902				& 0.9893		& $<${\bf 0.0001}	& 							& 						&				 \\
$a_p$			& 0.4458					& 0.5652					& 0.5036				& 0.9258						& 0.4294					& 0.4570				& 0.8521		& {\bf 0.0303}		& 0.2162				& 						& 				\\
$e_p$			& 0.2682					& 0.3735					& 0.3691				& 0.2266						& 0.5929					& 0.3681				& 0.3979		& 0.0645					& {\bf 0.0348}	& 0.2426			& 				\\
$J_p$			& 0.2301					& 0.1979					& 0.7299				& 0.2168						& 0.4060					& 0.3170				& 0.1022		& $<${\bf 0.0001} & {\bf 0.0029}	& 0.7167			& 0.4335\\   
\hline
$\log g$ 		& -0.6992				& 								& 							& 									& 								& 							& 					& 								& 							& 						&				\\
$[Fe/H]$		&				& 								& 							& 									& 								& 							& 					& 								& 							& 						&				\\
$V \sin i $	& 0.5506					& -0.4537				& 							& 									& 								& 							& 					& 								& 							& 						& 				\\
$R_{*}$		& 0.7509					& -0.9084				& 							& 									& 								& 							& 					& 								& 							& 						& 				\\
$M_{*}$		& 	0.9036					&	-0.8577				& 							&	0.5081						& 0.9407					& 							& 					& 								& 							& 						& 				\\
$J_{*}$		& 0.5840					& -0.6319				&							& 0.9312						& 0.5408					& 0.6327				& 					& 								& 							& 						& 				\\
$M_{p}$		& 								& 								& 							& 									& 								&							& 					& 								& 							& 						&				 \\
$R_{p}$		& 								& 								& 							& 									& 								&							& 					& 0.7468					& 							& 						&				 \\
$a_p$			& 								& 								& 							& 									& 								&							& 					& -0.4521				& 							& 						& 				\\
$e_p$			&								& 								& 							& 									& 								&							& 					&								& -0.4418			& 						& 				\\
$J_p$			&								& 								& 							& 									& 								&							& 					& 0.67921				& 0.5927				& 			           & 				\\  
\hline
 \end{tabular}
 }
 \end{Huge}
 \end{minipage}
 \end{table*}
 
As a final test, we have calculated the Pearson correlation matrices \citep[also explained in][]{Dalgaard2008} for the systems with HMEs and LMEs. The results can be found in Table~\ref{tab:3} and Table~\ref{tab:4}. 	Since the correlation matrices are symmetrical we show only the lower diagonal of each, showing first the matrix with the p-values(with alpha $\sim 0.05$), followed by the matrix for the Pearson correlation coefficient (keeping only the significant correlations, marked in bold in the matrix of the p-values). 

Comparing the HMEs with LMEs, there are obvious correlations of the temperature with the mass an radius in both systems, which suggests that, in general, as the temperature increases, the mass and radius increase. This explains, therefore, the strong correlations in both systems of the temperature with the velocity of rotation and thus the angular momentum. Note that these results are consistent with the general relation we determined in paper~{\rm I} between $V \sin i$, $T_{eff}$ and $\log g$. This suggests that the more massive the star the higher its rotation. 

One difference between the two systems is that although there are no correlations of the temperature with the surface gravity in the HMEs, there is one anticorrelations in the LMEs.  Due to the smallness of our samples, it is difficult to make sense of this difference physically. However, we note that $\log g$ in the LMEs is also well correlated with the radius, the mass and angular momentum (but not $V \sin i$ itself), something that is not seen in the HMEs. In Fig.~\ref{f2} wee see that the dispersion of $V \sin i$  decreases at low temperature, which implies that the bi-exponential relation of $V \sin i$ as a function of $T_{eff}$ and $\log g$ becomes tighter, suggesting that the behavior of $\log g$ becomes more ordered, which could explain the anti correlations with $T_{eff}$ and correlations with $M_{*}$ and $R_{*}$ ( and thus also with $J_{*}$). 

It is remarkable to note that $[Fe/H]$ in both systems is not correlated with any of the other parameters, either related to the stars or the planets. For the stars, the most obvious correlations in both systems are between $M_{*}$ and $R_{*}$, or $V \sin i$ and $J_{*}$. Note also that although $V \sin i$ shows no correlation with $M_{*}$ and $R_{*}$ in the HMEs it is correlated with $M_{*}$ in the LMEs. This might also explain why $R_{*}$ is not correlated with $J_{*}$ in the HMEs, while it is in the LMEs. 

In the case of the planets, the most important result of this analysis is the (almost complete) absence of correlations between the parameters of the planets and the parameters of the stars (most obvious in the LMEs). The only parameter that show some correlations with the star parameters is the semi-axis, $a_p$, of the orbit of the planets in the HMEs. This might suggests a difference in terms of circularization. Two results also seem important. The first is that $R_p$ is only correlated to $M_p$ in in the HMEs, which is consistent with the fact that the radius of the HMEs is constant. The second is that there is no correlation between $J_{p}$ and $J_{*}$, which, consistent with  the behavior observed in Figure~\ref{f5}, and which could suggest there is no dynamical coupling between the two. This, probably, is due to important losses of angular momentum during the formation of the stars and migrations of the planets. 

\section{Discussion}
\label{discussion}

Although our sample is small, we do distinguish a connection between the exoplanets and their stars: massive exoplanets tend to form around more massive stars, these stars being hotter (thus brighter) and rotating faster than less massive stars.  When we compare the spin of the stars with the angular momentum of the orbits of the exoplanets, we found that in the HME systems both the stars and planets rotate faster than in the LME systems. This is consistent with the idea that massive stars formed more massive PPDs, which rotate faster, explaining why the planets forming in these PPDs are seen also to rotate faster. 

\begin{figure}
\includegraphics[width=0.7\linewidth, angle=270]{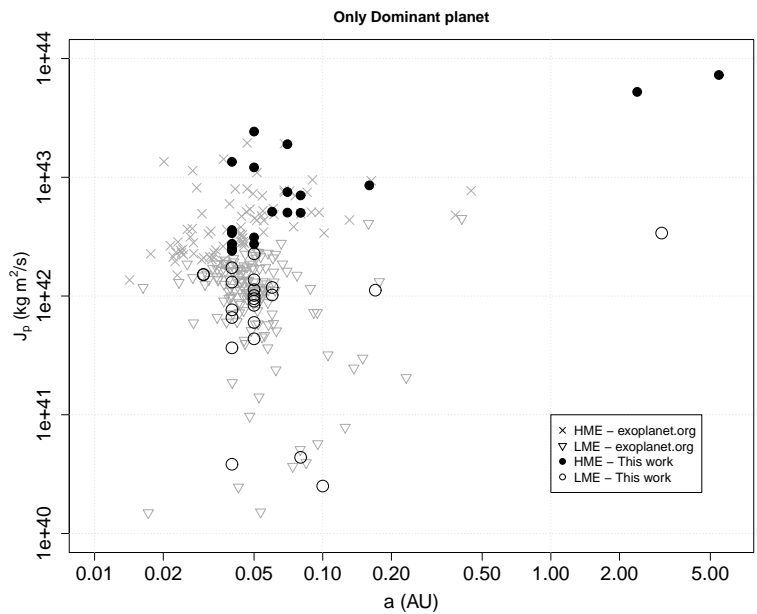}
\caption{\textit{Angular momentum of the exoplanets in our sample with respect to their distances from their stars. For comparison the exoplanets in our initial sample are also shown in light gray.}}
\label{f6}
\end{figure}

When we compare the effective angular momentum of the stars (Figure~\ref{f5}a) we found no evidence that they follow a $j-M$ relation, in particular, like the one suggested by \citet{McNally1965}, or that the angular momentum of the systems (Figure~\ref{f5}b) follow such relation. Furthermore, there is correlation between $J_{p}$ and $J_{*}$, which suggests that there is no dynamical coupling between the two. This is probably due to the important losses of angular momentum of the stars during its formation (by a factor $10^6$) and of the planets during their migrations (higher than 80\% their possible initial values). For the planets, their final angular momentum depend on their masses where its migration ends, $a_p$. Assuming, consequently that they all form more or less at the same distance (farther than the ice line in their systems) and end their migration at the same distance from their stars (which seems to be the case, as we shown in Figure~\ref{f6}), the HMEs would have lost a slightly higher amount of angular momentum than the LMEs. Within the scenario suggested above, this might suggest that more massive PPDs are more efficient in dissipating the angular momentum of their planets.

One thing seems difficult to understand, however, which is considering HMEs are more massive and lost more angular momentum during their migration, why did  they end their migration at almost exactly the same distance from their stars as the less massive LMEs. In Fig.~\ref{f6} the accumulation we perceive between 0.04 and 0.05 AU is consistent with the well known phenomenon called the three-days pile-up (3 days is equivalent to slightly higher than 0.04 AU, assuming Kepler orbits), which is supposed to be an artifact due to selection effects of ground-based transit surveys \citep{Gaudi2005}. However, \citet{Dawson2018} in their review about migration suggested this could be physical, and some authors did propose different physical explanations \citep[see][and references therein]{Fleck2008}. The model of \citet{Fleck2008} is particularly interesting because it tries to solve the problem using the same structure of the PPD that many authors believe explains how the stars loose their angular momentum during the T-Tauri phase, that is, magnetic braking. In the Appendix~\ref{sec:ap-B} we did some calculations which show that where the planets in our sample end their migration could be close to the co-rotation radius, the region of the PPD where the disk turns at the same velocity as the star. But does this imply that disk migration is more probable than high-eccentricity tidal migration?

\begin{figure}
\includegraphics[width=0.9\linewidth]{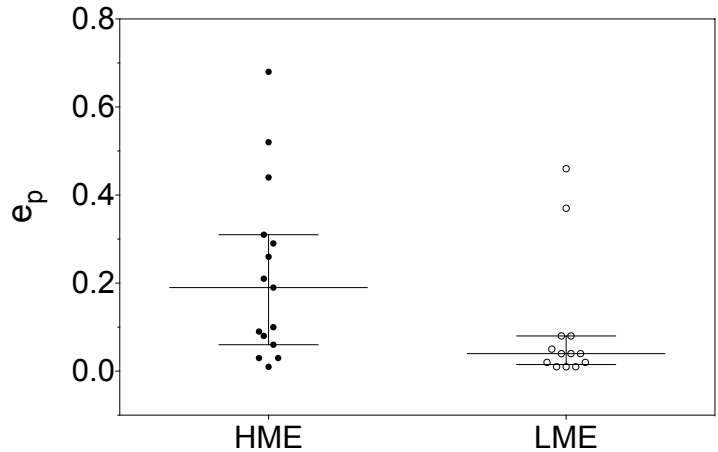}
\caption{\textit{Eccentricities of the exoplanets in our sample, distinguishing between HMEs and LMEs.}}
\label{f7}
\end{figure}

According to the theory of high-eccentricity tidal migration,  one expects a strong dependence of the tidal evolution timescale on the final location of the orbit of the planets, $a_{final}$ \citep[e.g.,][]{Eggleton1998}: 
\begin{equation}
\dot{a} \propto a_{final}^8
\end{equation}
This implies that since the HMEs and LMEs have the same $a_{final} \sim 0.04$ AU they should also have the same tidal evolution timescale, and thus no difference would be expected comparing their eccentricities. What stops the planet migration in the high-eccentricity tidal migration model is the circularization of the orbit through tidal interactions with the central stars. Assuming same tidal evolution timescale, therefore, we would expect the eccentricities for the HMEs and LMEs to be all close to zero \citep{Bolmont2011,Remus2012}. Note that using our data to test whether the HMEs and LMEs have similar distributions in eccentricity is possibly difficult, because we are not certain an eccentricity of zero is physical or not (meaning an absence of data). For the HME sample, on 22 exoplanets we count 7 (32\%) with zero eccentricity, while in the LMEs out of 23, 10 (43\%) have zero eccentricity. The difference seems marginal. For the remaining planets with non zero eccentricities (15 HMEs and 13 LMEs) we compare in Fig.~\ref{f7} their distributions. There is a weak difference, with a median (mean) of $e_p = 0.19$ ($e_p = 0.22$) for the HMEs compared, to $e_p = 0.04$ ($e_p = 0.10$) for the HMEs and LMEs. A MW test yields a p-value = 0.0257, which suggests a difference at the lowest significance level. Therefore, there seems to be a trend for the HMEs to have on average a higher eccentricity than the LMEs. The HMEs possibly reacted more slowly to circularization than the LMEs \citep[][]{deWit2016}, suggesting possibly different structures due to their higher masses \citep{Flor-Torres2016}.

\section{Conclusions}
\label{conclusion}

Based on an homogeneous sample of 46 stars observed with TIGRE and analysed using \textsf{iSpec} we started a project to better understand the connection between the formation of the stars and their planets. Our main goal is to check is there could be a coupling between the angular momentum of the planets and their host stars. Here are our conclusions. 

There is a connection between the stars and their exoplanets, which passes by their PPDs. Massive stars rotating faster than low-mass stars, had more massive PPDs with higher angular momentum, explaining why they formed more massive planets rotating faster around their stars. However, in terms of stellar spins and planets orbit angular momentum, we find that both the stars and their planets have lost a huge amount of angular momentum (by more than 80\% in the case of the planets), a  phenomenon which could have possibly erased any correlations expected between the two. The fact that all the planets in our sample stop their migration at the same distance from their stars irrespective of their masses, might favor the views that the process of migration is due to the interactions of the planets with their PPDs and that massive PPDs dissipates more angular momentum than lower mass PPD. Consistent with this last conclusion, HMEs might have different structures than LMEs which made them more resilient to circularization.  

\acknowledgments

We like to thank an anonymous referee for his careful revision of our results and for his comments and suggestions that helped us improved our work. L. M. F. T. thanks S. Blanco-Cuaresma for discussions and support with \textsf{iSpec}. She also thanks CONACyT for a scholar grant (CVU 555458) and travel support (bilateral Conacyt-DFG projects 192334, 207772, and 278156), as well as supports given by the University of Guanajuato for conference participation and international collaborations (DAIP, and Campus Guanajuato). This research has made use of the Exoplanet Orbit Database, the Exoplanet Data Explorer \citep[exoplanets.org][]{Han2014}, exoplanets.eu \citep{Schneider2011} and the NASA's Astro-physics Data System.

\begin{appendices}
    \section{Calculating the sefl-gravitating mass limit}
    \label{sec:ap-A}
According to Padmanabhan (1993) the formation of structures with different masses and sizes involves a balance between two forces, gravity, $F_g$ and electromagnetic, $F_e$. For the interaction between two protons, we get:
 \begin{equation}
 \frac{F_e}{F_g}=\frac{\kappa_e e^2/r^2}{Gm_p^2/r^2}=\frac{\kappa_e e^2}{Gm_p^2}=\left( \frac{\kappa_e e^2}{\hbar c}\right) \left( \frac{\hbar c}{Gm_p^2}\right)
 \end{equation}
 where $\kappa_e$ and $G$ are the electromagnetic and gravitational constants, that fix the intensity of the forces, $e$ and $m_p$ the charges and masses interacting ($m_p$ is the mass of a proton) and $r$ the distance between the sources (the laws have exactly the same mathematical form). Introducing the reduced Plank constant $\hbar = h/2\pi$ and the velocity of light $c$ (two important constant in physics), the first term on the right is the fine structure constant $\alpha \sim 7.29 \times 10^{-3}$, and the second term is the equivalent for gravity $\alpha_G \sim5.88 \times 10^{-39}$. This implies that:
 \begin{equation}
 \frac{\alpha}{\alpha_G} \sim 1.24 \times 10^{36}
\end{equation}  

This results leads to the well-known hierarchical problem: there is no consensus in physics why the electromagnetic force should be stronger than the gravitational force in such extreme. The reason might have to do with the sources of the forces, in particular, the fact that in electromagnetic there are two types of charge interacting, while there is only one type of mass. The important consequence for the formation of large-scale structures is that while in the electromagnetic interaction the trend to minimize the potential energy reduces the total charge, massive object being neutral, the same trend in gravity is for the gravitational field to increase with the mass. Therefore, small structures tend to be dominated by the electromagnetic force, while large structures are dominated by gravity. Based on this physical fact, Padmanabhan calculated that there is a critical mass, $M_c$, above which the force of gravity becomes more important than the electromagnetic force. The point of importance for planet formation is that this critical mass turned out to be comparable to the mass of a gas giant planet. 

To realize that, it suffices to compare the energy of ligation of a structure with its gravitational potential energy. Since the escape energy for an electron is:
\begin{equation}
E_0 \approx \alpha^2 m_e c^2 \simeq 4.35 \times 10^{-18} {\rm J}
\end{equation}  
a spherical body formed of $N$ atoms would have an energy of ligation $E = N \times E_0$. On the other hand, its mass would be $M = N m_p$ and its volume would be $4/3 \pi R^3 = N \times 4/3\pi a_0^3$, where $a_0 \simeq 5.30 \times 10^{-11}$ m, is the Bohr radius. From this we deduce that its radius woud be $R \sim N^{1/3} a_0$, and its gravitational potential:
\begin{equation}
E_G \simeq \frac{GM^2}{R}\simeq N^{5/3}\frac{Gm_p^2}{a_0}=N^{5/3}\alpha_G m_e \alpha c^2
\end{equation}
Now, the condition for a stable object to form under gravity is $E \geq E_G$, while for $E < E_G$ the object would collapse under its own weight. In equilibrium we would thus have:
\begin{equation}
N^{5/3}\alpha_G m_e \alpha c^2 = N \alpha^2 m_e c^2
\end{equation}
This yields to a maximum number of atoms,  $N_{max} \sim (\alpha/\alpha_G)^{3/2} \sim 1.38 \times 10^{54}$, and a critical mass:
\begin{equation}
M_c = N_{max} \times m_p \sim 2.31 \times 10^{27} {\rm kg} \sim 1.2 M_J
\end{equation}
This suggests that the inflection point we observe in the mass-radius relation of exoplanets could be the critical point where an object becomes unstable under self-gravity. 
    

\section{At what distance is the corotation radius, $R_{co}$}
    \label{sec:ap-B}

In their review about the migration of planets, \citet{Dawson2018} claimed that the distribution of exoplanets around their host stars is consistent with the co-rotation radius, $R_{co}$, which is part of the magnetic structure connecting the star to its PPD. In the model presented by \citet{Matt2004}, the authors discussed some specific aspects of this magnetic structure that could contribute in stopping planet migration (see their Figure~3 and explanations therein). Following this model, the star and disk rotate at different angular speeds except at $R_{co}$, where the magnetic field becomes twisted azimuthally by differential rotation, triggering magnetic forces. These forces would act to restore the dipole configuration conveying torques between the star and the disk. As a planet approaches $R_{co}$, therefore, these torques would transfer angular momentum from the star to the planet stopping its migration \citep{Fleck2008}. One important aspect of this model is that the dumping would only work up to a distance $R_{out} \sim 1.6 R_{co}$, and thus one would expect the planets to pile-up over this region, that is, between $R_{out}$ and $R_{co}$. The question then is what is the value of $R_{co}$?
  
One way to determine this value is to assume that when the wind of a newly formed star starts evaporating the PPD,  $R_{out} \longrightarrow R_{co}$ and the magnetic pressure at $R_{co}$ balances the gas pressure due to the wind, $P_B = P_g$. Assuming, that the  intensity of the magnetic field decreases as the cube of the distance, we get, $B \sim B_s R_{*}^3/r^3$, where $B_s$ is the intensity of the magnetic field at the surface of the star, $R_{*}$ is the radius of the star, and $\mu_0$ is the permeability of the vacuum. This yields that: 
\begin{equation}\label{PB}
P_B = \frac{B^2}{2 \mu_0} = \frac{B^2_s R_{*}^6}{2 \mu_0 r^6}
\end{equation}
For the gas pressure we used the expression for ``Ram pressure'':
\begin{equation} 
P_g = n m v^2
\end{equation}
where $v$ is the velocity of the wind and $n m$ is its load, that is, the amount of mass transported by the wind. This load then can be expressed in terms of the flux of matter, $\dot{M}$, as:
\begin{equation}
n m = \frac{\dot{M}}{4 \pi r^2 v}
\end{equation}
Assuming equality at $r = R_{co}$, we get:
\begin{equation}
R^4_{co} = \frac{2\pi}{\mu_0} \frac{B^2_s R^6_{*}}{\dot{M} v}
\end{equation}
which for a star like the Sun ($B_s \sim 10^{-4}$\ T, $\dot{M} \sim 2.00 \times 10^{-14} M_\odot/\rm{yr}$ and $v = 2.15 \times 10^{-3}$\ m/s) yields $R_{co} \sim 6.77 \times 10^9 \rm{m}$ or  $ 0.045$ AU. According to this model, therefore, we could expect to see a real pile up of exoplanets independent of their masses (and losses of angular momentum) between $0.04 - 0.07$ AU, which neatly fit the observations. 
  \end{appendices}
  

\end{document}